\documentstyle[prl,aps,epsf]{revtex}
\begin{document}
\draft
\title{A new approach to the ground state of quantum-Hall systems}
\author{S. A. Mikhailov}
\address{Institute for Theoretical Physics, University of Regensburg, 93040 Regensburg, Germany} 
\date{\today}
\maketitle
\begin{abstract}
I present variational solutions of the many-body Schr\"odinger equation for a two-dimensional electron system in strong magnetic fields, which have, at $\nu=1$, the energy about 46\% 
lower than the energy of the Laughlin liquid. At $\nu=2/3$, I obtain the energy, about 29\% 
lower than was reported so far.
\end{abstract} 

The need to explain the integer \cite{Klitzing80} and the fractional \cite{Tsui82} quantum Hall effects \cite{DasSarma97} (QHE) resulted in tremendous theoretical efforts \cite{Fukuyama79,Yoshioka79,Yoshioka83b,Maki83,Yoshioka83a,Laughlin83,Haldane83,Yoshioka84b,Lam84,Levesque84,Haldane85,MacDonald85,Kivelson86,Morf86,Jain89a,Halperin93,Bychkov81,Koulakov96} to understand the nature of the ground state of a system of strongly interacting two-dimensional (2D) electrons in strong magnetic fields. The modern, widely accepted explanation of the fractional QHE is based on the theory of Laughlin \cite{Laughlin83}, who showed that a special quantum state -- an incompressible quantum liquid with fractionally charged quasiparticles \cite{Goldman95,Goldman96,Goldman97,dePicciotto97,Saminadayar97,Reznikov99} -- has the lowest ground state energy at the Landau-level filling factors $\nu=1/m$, $m=1,3,5$. In this Letter I present an infinite set of variational many-body wave functions for a 2D electron system (ES) in strong magnetic fields ${\bf B}$, and an exact analytical method of calculating the energy of the new states. The new wave functions describe the system at {\it all} Landau-level filling factors $\nu\le 1$, {\it continuously as a function of magnetic field}. In certain intervals of $\nu$ (approximately, at $\nu\gtrsim 1/2$) the calculated upper limits to the ground-state energy are {\it substantially lower} than was reported up till now \cite{Laughlin83,Levesque84,Morf86}. 

The main theoretical efforts so far have been concentrated on the region of the very strong ($\nu<1$) magnetic fields \cite{Fukuyama79,Yoshioka79,Yoshioka83b,Maki83,Yoshioka83a,Laughlin83,Haldane83,Yoshioka84b,Lam84,Levesque84,Haldane85,MacDonald85,Kivelson86,Morf86,Jain89a,Halperin93}. The nature of the ground state of a completely filled lowest Landau level ($\nu=1$) seemed to be well understood: the Hartree-Fock many-body wave function 

\begin{equation}
\Psi_{HF}= \frac 1{\sqrt{N!}}\det|\psi_{L_j}({\bf r}_i)|,{\ \ }L_j=0,1,\dots,N-1,
\label{HF}
\end{equation}
-- a Slater determinant built from the single-particle lowest-Landau-level states 
\begin{equation}
\psi_L({\bf r})= \frac{(z^*)^L}{\lambda\sqrt{\pi L!}}\exp(-zz^*/2),
\end{equation}
-- was proposed about 20 years ago \cite{Bychkov81} and was considered to be the cornerstone of the theory [here $z=(x+iy)/\lambda$ is a complex coordinate of an electron, $\lambda^2=2\hbar c/eB$, $L=0,1,2,\dots$ is the angular momentum quantum number, and $\hbar$, $c$, and $e$ are the Planck constant, velocity of light and the electron charge, respectively]. The state (\ref{HF}) is a special case of the Laughlin liquid. It is characterized by the uniform electron density $n_s=1/\pi\lambda^2$ and the energy per particle \cite{Laughlin83}
\begin{equation}
\epsilon_{HF}=-\pi/2=-1.5708
\label{HFenergy}
\end{equation}
in the $B$-independent energy units $e^2\sqrt{n_s}$.

The wave function (\ref{HF}) describes electrons rotating around the {\it same} center (arbitrarily placed at the origin) with {\it different} angular momenta. However, in an infinite and uniform 2DES all points of the 2D plane are equivalent and no one of them can be considered as a special point, which electrons would prefer to rotate around. I consider, as trial functions for the ground state of the 2DES, an infinite set of many-body wave functions 
\begin{equation}
\Psi_L=\frac 1{\sqrt{N!}}\det|\chi^L_{ij}|,
\label{PsiL}
\end{equation}
\begin{equation}
\chi_{ij}^L\equiv\chi_L({\bf r}_i,{\bf R}_j)=\psi_{L}({\bf r}_i-{\bf R}_j) e^{-i\pi{\bf r}_i\cdot({\bf B\times R}_j)/\phi_0},
\label{chi}
\end{equation}
which describe a system of electrons rotating around {\it different} points ${\bf R}_j$ with the {\it same} angular momentum. Here $\phi_0=hc/e$ is the flux quantum, and $L=0,1,2,\dots$. The points ${\bf R}_j$ are chosen in view of minimizing the Coulomb energy of the system, and in a pure 2DES (without disorder) coincide with points of a triangular (or a square) lattice, uniformly distributed over the 2D plane with the average density $n_s$ ($n_sa^2=2/\sqrt{3}$ and 1 for the triangular and the square lattice, respectively, $a$ is the lattice constant). All the states $\Psi_L$ are the eigen functions of the kinetic energy operator, with the energy $\hbar\omega_c/2$ and the angular momentum $L$ per particle. The state $\Psi_{L=0}$ is the Wigner crystal state \cite{Maki83}. The states $\Psi_{L>0}$ have been discussed in the literature \cite{Fertig99}, but only in Hartree approximation, when the neighbour single-particle states $\psi_{L}$ do not overlap. As will be seen below, the most interesting physics takes place just in the opposite case of a {\it strong overlap} of the states $\psi_{L}$ centered at different lattice points ${\bf R}_j$. 

In order to calculate expectation values of different physical parameters with the wave functions (\ref{PsiL}), I derive exact analytical formulas for the norm 
\begin{equation}
\langle\Psi_L|\Psi_{L}\rangle=\det|{\cal S}|, 
\label{norm}
\end{equation}
and matrix elements of arbitrary single-particle, $\langle\Psi_L|\hat H_1|\Psi_{L}\rangle$, and two-particle, $\langle\Psi_L|\hat H_2|\Psi_{L}\rangle$, operators. The norm (\ref{norm}) and the matrix element of Coulomb energy,
\begin{equation}
\langle\Psi_L|V_C|\Psi_{L}\rangle=\frac 12
\sum_{ijkl}(-1)^{i+j+k+l}
{\rm sgn}(i-k)
 {\rm sgn}(j-l)V_{ijkl}^{LL}\det |{\cal S}|^{row\neq i,k}_{column\neq j,l},
\label{coulomb}
\end{equation}
are expressed via the matrix ${\cal S}$ with the elements
\begin{equation}
S_{ij}\equiv S_{ij}^{LL}=\langle
\chi_{L}({\bf r},{\bf R}_i)|\chi_{L}({\bf r},{\bf R}_j)\rangle,
\end{equation}
and the 
Coulomb matrix elements $V_{ijkl}^{LL}=\langle\chi_{ai}^L\chi_{bk}^L| e^2|{\bf r}_a-{\bf r}_b|^{-1}|\chi_{aj}^{L}\chi_{bl}^{L}\rangle$. The sums in (\ref{coulomb}) are taken over $N\gg 1$ lattice points. Both the overlap integrals $S_{ij}^{LL}$ and the matrix elements $V_{ijkl}^{LL}$ can be analytically calculated, so that I get, finally, {\it exact closed-form analytical expressions} for the energy $\epsilon_L$ of the system (per particle) in the states $\Psi_L$. The final result can be presented in the form $\epsilon_L=\epsilon_L^{H}+\epsilon_L^{xc}$. The Hartree contribution $\epsilon_L^{H}$ contains the sum $\sum_{i<j} V_{iijj}^{LL}$ and the energy of the uniform positive background. The exchange-correlation energy contains two-site exchange terms $V_{ijji}$, as well as three-site ($V_{iikl}$, $V_{ijki}$) and four-site ($V_{ijkl}$) correlations. 

\begin{figure}%[ht]
\begin{center}
\begin{math}
\epsfxsize=8.5cm
\epsffile{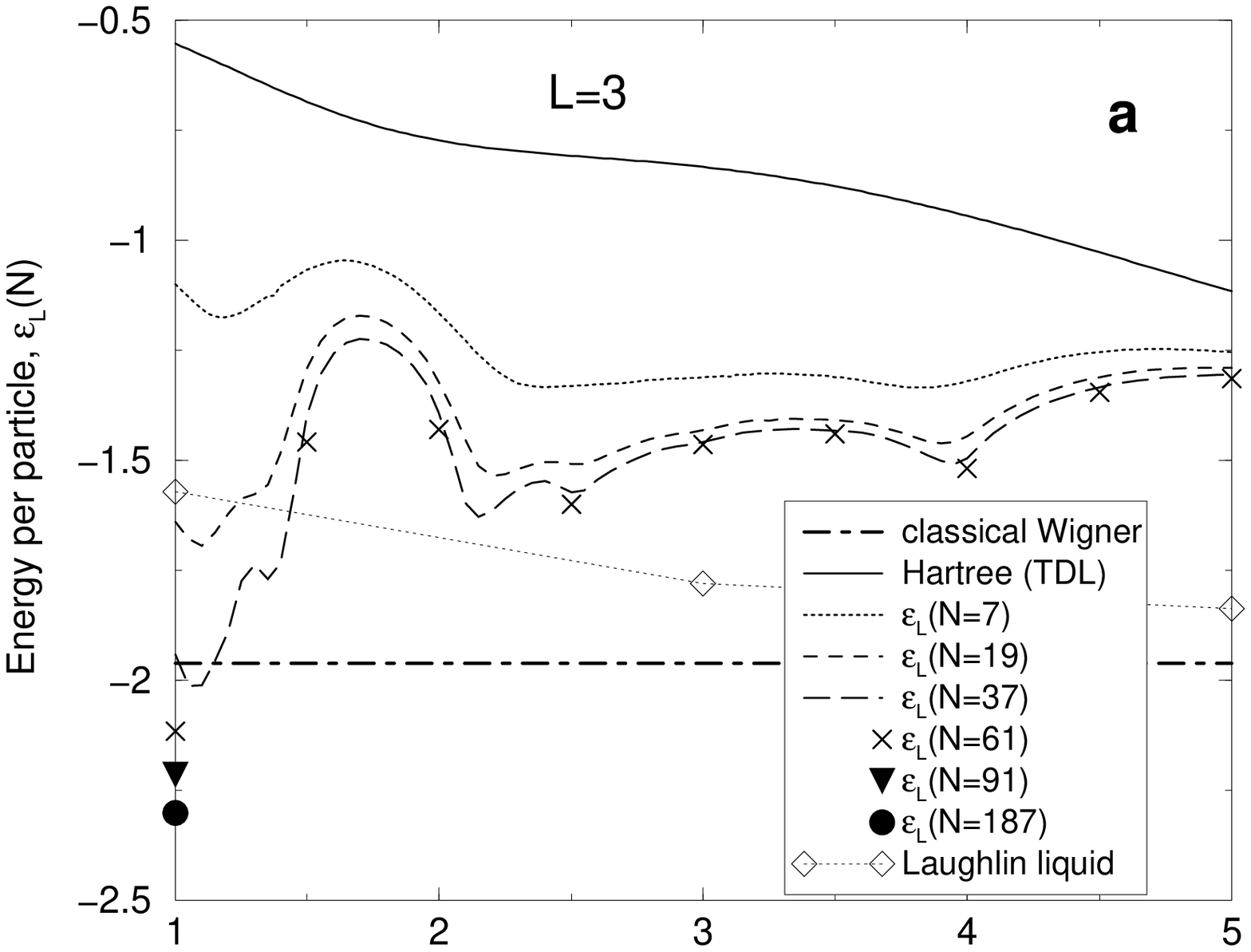}
\end{math}\\
\begin{math}
\epsfxsize=8.5cm
\epsffile{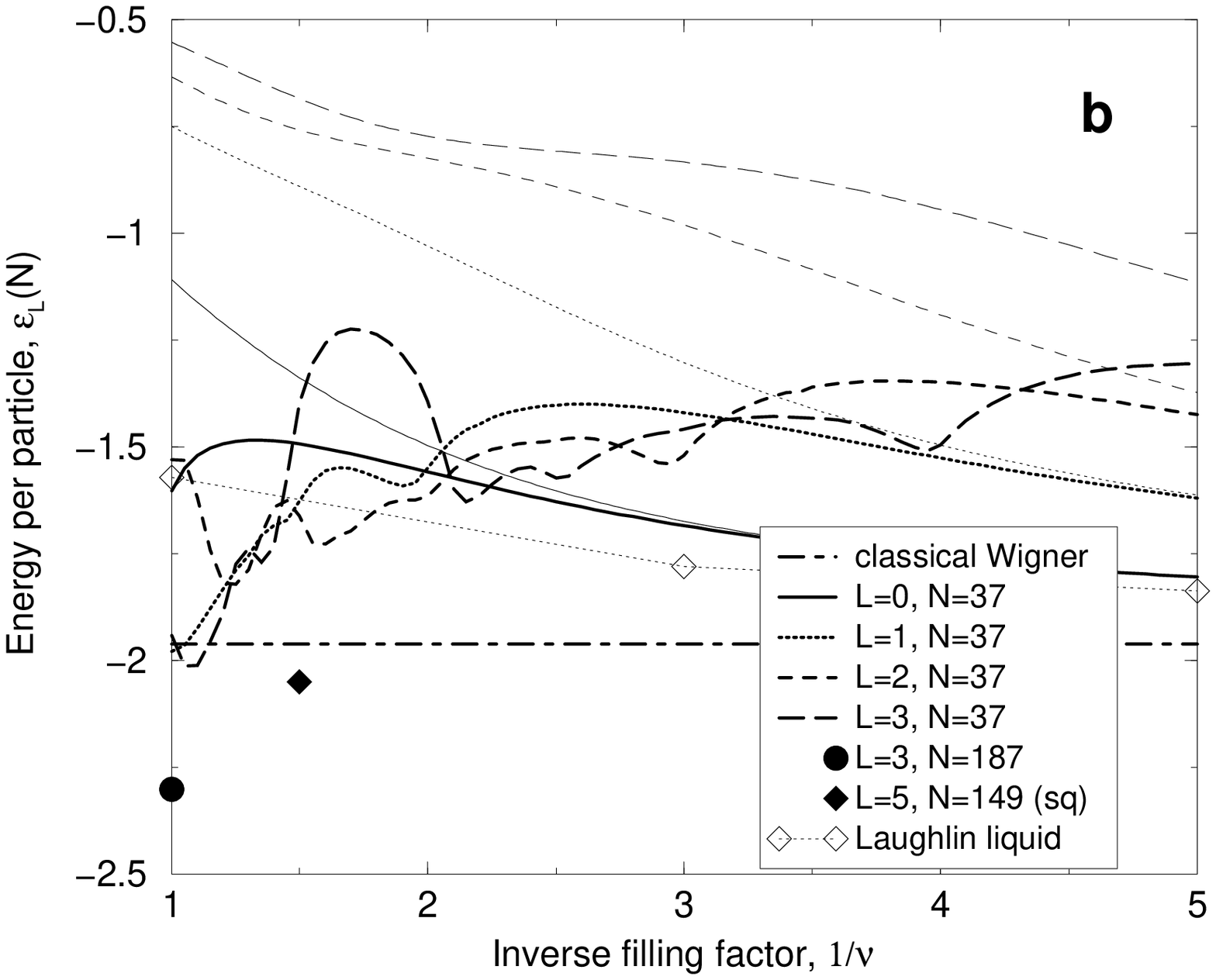}
\end{math}
\end{center}
\caption{Calculated energies $\epsilon_L(N)=\epsilon_L^H+\epsilon_L^{xc}(N)$ 
per particle (in units $e^2\sqrt{n_s}$) as a function of magnetic field (of the inverse filling factor $\nu^{-1}=B/n_s\phi_0$): {\bf a}, at $L=3$ and different numbers of lattice points $N$; {\bf b}, in the states $\Psi_L$ with a few lowest $L$. The filled diamond in {\bf b} corresponds to a square lattice of points ${\bf R}_j$, all other points and curves -- to a triangular lattice. The Hartree contribution $\epsilon_L^H$ (the uppermost curve in {\bf a}, thin curves in {\bf b}) is calculated exactly in the thermodynamic limit $N=\infty$. For comparison, the energy of the classical Wigner crystal \protect\cite{Bonsall77} and of the Laughlin liquid \protect\cite{Laughlin83,Levesque84} are also shown.}
\label{fig1}
\end{figure}

\begin{figure}%[t]
\begin{center}
\begin{math}
\epsfxsize=8.5cm
\epsffile{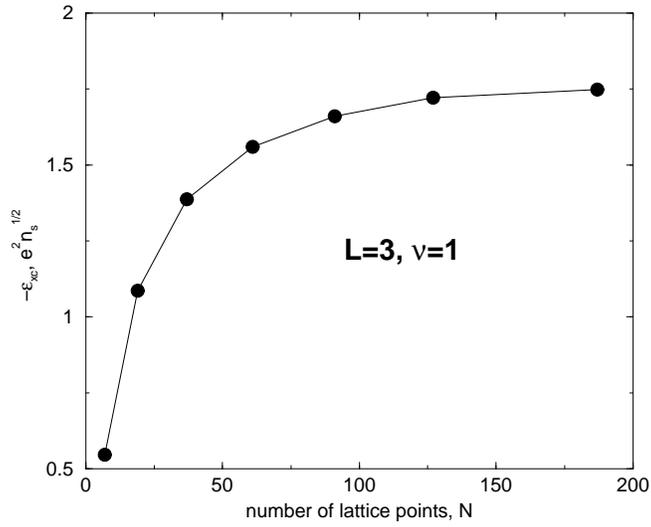}
\end{math}
\end{center}
\caption{Absolute value of the exchange-correlation energy per particle, $-\epsilon_L^{xc}(N)$, in units $e^2\sqrt{n_s}$, as a function of the number of (triangular) lattice points $N$, for the state $\Psi_{L=3}$ at $\nu=1$.}
\label{fig-xc}
\end{figure}

The number $N$ of lattice points involved in calculation of sums (\ref{coulomb}) should, ideally, be infinite. I calculate the Hartree energy $\epsilon_L^{H}$ exactly, in the thermodynamic limit $N=\infty$, and the exchange-correlation energy $\epsilon_L^{xc}(N)$ approximately, at a finite number of lattice points. It turns out that the contribution $\epsilon_L^{xc}(N)$ is {\it negative} and grows with $N$ in its absolute value, Figures \ref{fig1}a and \ref{fig-xc}. The calculated energy $\epsilon_L(N)$ is thus an upper limit to the true energy $\epsilon_L$, $\epsilon_L<\epsilon_L(N)$. 

Figure \ref{fig1} exhibits $B$-dependencies of the calculated energies $\epsilon_L(N)$ at a number of different $L$ and $N$ in a system with a triangular lattice of points ${\bf R}_j$. As expected, in strong magnetic fields the states with larger $L$ have larger energy, due to a larger Hartree contribution. When $B$ decreases, however, the single-particle wave functions rotating around different lattice points begin to overlap, and exchange-correlation interaction significantly reduces the energy. Due to the overlap of electron wave functions and electron interference effects, the energy of the system in the states $\Psi_L$ has a complex oscillating dependence on magnetic field. As a consequence, in different intervals of $B$ the ground state of the system is realized by states with different $L$. The gap between the ground and the first excited state also varies with $B$ and vanishes in points, where two different $L$-states have the same energy. It seems likely that, maxima in the diagonal resistivity of the system are observed in these, zero-gap, points, while the Hall-resistivity plateaus are seen in the regions of large gaps, when the system is in the same quantum state. This view on the integer and the fractional QHEs merits detailed consideration and requires further development of the theory. 

At $\nu=1$, the state $\Psi_{L=3}$ has the lowest energy. For this state I calculate $\epsilon_{L=3}^{xc}(N)$ with inclusion of up to $N=187$ lattice points, Figures \ref{fig1} and \ref{fig-xc}. Thus calculated upper limit to the total energy per particle, 
\begin{equation}
\epsilon_{L=3}(\nu=1)<\epsilon_{L=3,N=187}(\nu=1)=-2.302,
\label{nu1}
\end{equation}
is more than 46\% 
lower than the energy (\ref{HFenergy}) of the Hartree-Fock state (\ref{HF}). 
Note that the energy (\ref{nu1}) is even lower than the energy of the classical Wigner crystal \cite{Bonsall77}.

\begin{figure}%[h]
\begin{center}
\begin{math}
\epsfxsize=8.5cm
\epsffile{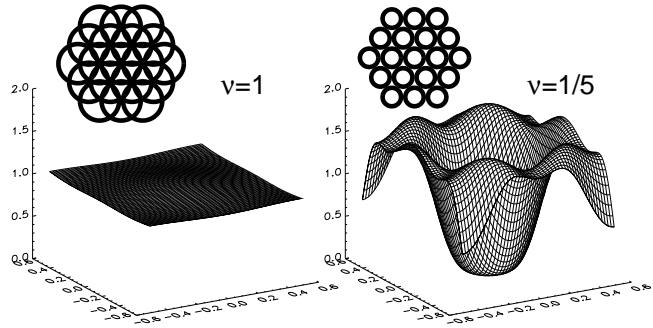}
\end{math}
\end{center}
\caption{Calculated normalized density of electrons $n_e({\bf r})/n_s$ in the state $\Psi_{L=3}$ at $\nu=1$ (left panel) and $\nu=1/5$ (right panel). Up to $N=301$ lattice points have been included in this calculation. Insets shows an overlap of electron rings centered at sites of the triangular lattice ${\bf R}_j$.}
\label{fig2}
\end{figure}

The state $\Psi_{L=3}$ at $\nu=1$ is characterized by a very strong overlap of the single-particle wave functions of neighbour electrons, Figure \ref{fig2}. As a consequence, the density of electrons $n_e({\bf r})$ in this state is uniform (left panel). The wave function $\Psi_{L=3}(\nu=1)$ thus describes a quantum liquid with strong exchange-correlation interaction of a very large number of electrons. [For comparison, the right panel shows the electron density in the same state ($L=3$) at $\nu=1/5$ (the state $\Psi_{L=3}$ is one of the excited states at $\nu=1/5$, Figure \ref{fig1}b). Qualitative difference between the liquid-type state at $\nu=1$ and a solid-type state at $\nu=1/5$ is obvious]. 

The wave functions $\Psi_L$ have a very large variational freedom. Apart from a possibility to vary the angular momentum $L$, one can consider configurations with a different (e.g. square) symmetry of lattice points ${\bf R}_j$. For instance, I have observed that, at $\nu=2/3$ the {\it square} lattice configuration gives the energy, {\it lower} than the triangular one. The upper limit to the total energy per particle at $\nu=2/3$, calculated with $L=5$ and $N=149$ in the square lattice configuration (the filled diamond in Figure \ref{fig1}b), 

\begin{equation}
\epsilon_{L=5}(\nu=2/3)<\epsilon_{L=5,N=149}(\nu=2/3)=-2.050,
\label{nu23}
\end{equation}
is about 29\% 
lower than was reported so far \cite{Morf86} and also lower than the energy of the classical Wigner crystal. 

Can the energy of the states $\Psi_L$ be comparable with or lower than that of the Laughlin liquid at $\nu=1/3$ and $\nu= 1/5$? This important question remains open. Calculations performed so far for $L\le 3$ and $N\leq 91$ gave the energy higher than that of the Laughlin liquid at these values of $\nu$. However, Figure \ref{fig1} demonstrates two important tendencies: i) the downward cusps in dependencies $\epsilon_L(B)$ are being shifted to the right with growing $L$; and ii) for given $L$ and $\nu$ the energy $\epsilon_L(N)$ substantially decreases with the growth of $N$. Calculations with other $L$ and larger $N$ may lead to a reduction of energy $\epsilon_L$ below the energy of the Wigner-crystal state $\epsilon_{L=0}$ also at $\nu\le 1/3$. 

The new approach presented here enables one to analytically calculate any physical value characterizing the system {\it at all $\nu$, continuously as a function of magnetic field}. It naturally describes a transition to the Wigner crystal state at $\nu\to 0$, and can be straightforwardly generalized to the case of non-spin-polarized systems and hence to the region of higher Landau levels $\nu>1$. It also makes possible to study the interplay between disorder and electron-electron correlations in the integer and the fractional QHE (by varying configurations of points ${\bf R}_j$ which are influenced by disorder in a real system \cite{Cha94}). Due to a large variational freedom of wave functions $\Psi_L$ it opens up wide possibilities to search for better approximations to the ground state at all values of $\nu$.

This work was supported by the Max Planck Society and the Graduiertenkolleg {\sl Komplexit\"at in Festk\"orpern}, University of Regensburg, Germany. I also thank Peter Fulde and Ulrich R\"ossler for support of this work, Nadejda Savostianova for numerous helpful discussions, and Vladimir Volkov for useful comments.

\bibliography{fqhe}
\bibliographystyle{prsty}

\end{document}